\documentclass[aps,pra,10pt,reprint,superscriptaddress]{revtex4-2}
\usepackage{amsmath,amsfonts,amssymb}
\usepackage{xcolor,graphicx}
\usepackage{braket}
\usepackage[pdftex]{hyperref}
\usepackage{csquotes}
\hypersetup{
	colorlinks = true,
	linkcolor = blue,
	anchorcolor = blue,
	citecolor = red,
	filecolor = blue,
	urlcolor = blue}
\begin{document}
\title{Unified approach to paraxial propagation in uniform media and media with linear or quadratic refractive index distribution}

\author{N. Korneev}
\email[e-mail:\,]{korneev@inaoep.mx}
\affiliation{Instituto Nacional de Astrofísica Óptica y Electrónica, Calle Luis Enrique Erro No. 1\\ Santa María Tonantzintla, Puebla, 72840, Mexico}

\author{I. Ramos-Prieto}
%\email[e-mail:\,]{iran@inaoep.mx}
\affiliation{Instituto Nacional de Astrofísica Óptica y Electrónica, Calle Luis Enrique Erro No. 1\\ Santa María Tonantzintla, Puebla, 72840, Mexico}

\author{F. Soto-Eguibar}
\affiliation{Instituto Nacional de Astrofísica Óptica y Electrónica, Calle Luis Enrique Erro No. 1\\ Santa María Tonantzintla, Puebla, 72840, Mexico}

\author{U. Ru\'{i}z}
\affiliation{Instituto Nacional de Astrofísica Óptica y Electrónica, Calle Luis Enrique Erro No. 1\\ Santa María Tonantzintla, Puebla, 72840, Mexico}

\author{D. {S\'anchez}-{de-la-Llave}}
\affiliation{Instituto Nacional de Astrofísica Óptica y Electrónica, Calle Luis Enrique Erro No. 1\\ Santa María Tonantzintla, Puebla, 72840, Mexico}

\author{H. M. Moya-Cessa}
\affiliation{Instituto Nacional de Astrofísica Óptica y Electrónica, Calle Luis Enrique Erro No. 1\\ Santa María Tonantzintla, Puebla, 72840, Mexico}

\date{\today}

\begin{abstract}
We report explicit equations and matrix representations that allow simple calculation between three different media (free space, linear and quadratic refractive index distributions) for paraxial light propagation. 
\end{abstract}

\maketitle

\section{Introduction}
The paraxial wave equation describes light propagation within the paraxial approximation \cite{Grynberg2012,Kok2010,Boyd2008}. Mathematically, it is equivalent to the Schrödinger equation of quantum mechanics \cite{Stoler_1980,MoyaCessa2024,Ramos_2024}. It is well-established that there exists a simple relationship between solutions of the paraxial equation that initially differ by a spatial shift or by multiplication with an exponential factor $\exp(i\nu x)$~\cite{Torre_2010}. This mathematical relationship is associated with the existence of a certain Lie algebra that includes a Laplace operator, first derivative, multiplication by $x$, and unity. A less trivial consequence of this algebra is the ability to relate free propagation to propagation in a medium with a linear refractive index distribution.

In the context of quantum mechanics, it was realized as early as 1970 that another important Lie algebra exists \cite{Kalnins_1974a}, involving the Laplace operator along with a generator of the squeezing operator $x\partial_x$ and multiplication by $x^2$. This algebra is equivalent to the set of ABCD transformations \cite{Collins1970,Moshinsky1971,Ogura2009} and permits the relationship between solutions for wavefunctions that initially differ by multiplication with a phase factor $\exp(\frac{igx^{2}}{2})$ or by a Gaussian function $\exp(-\frac{gx^{2}}{2})$~\cite{Torre_2010}. Moreover, this algebra facilitates the direct relation between free space propagation and propagation in a medium with a quadratic refractive index distribution.

While many of the specific results discussed here are known in the mathematical, quantum mechanical, or optical (linear and nonlinear) literature, we believe that presenting these relations in an unified form with minimal use of abstract Lie algebraic terms can be beneficial for propagation studies. Firstly, this equivalence facilitates the adaptation of known solutions in free space, such as Hermite-Gauss and Laguerre-Gauss beams, Bessel beams, Airy beams, etc., to either linear or quadratic refractive index distributions, leading to a better qualitative understanding of the corresponding solutions~\cite{Durnin_1987,Gori_1987,Gutierrez_2000,Roux_2003,Bandres_2004,McGloin_2005,Gutierrez_2005,Bandres_2007,Khonina_2011,Ismail_2012,Sosa_2017,Cabrera_2017}. Additionally, one can express ABCD-type relations, which describe the propagation of beams with quadratic phase or amplitude modulation, in an unified manner for all three types of media based on the known propagation of a non-modulated beam. Secondly, numerical calculations of propagation are often easier to perform in free space, where the fast Fourier transform is highly effective for long propagation distances. For modulated index media, some variant of split-step calculation is usually necessary, which can pose problems of stability and computation time.

The paper is organized as follows: Sec.~\ref{Sec2} provides fundamental insights into Lie groups and algebras, establishing a framework for understanding linear operators that form a Lie algebra and their commutation relations. It outlines the basic properties of Lie algebras and sets the stage for analyzing paraxial light propagation in different media. Sec.~\ref{Sec3} focuses on the $SL(2,\mathbb{C})$ Lie algebra, describing its operators, commutation relations, and representation with $2\times2$ matrices. It explores the exponential of these matrices, elucidating the transformation properties relevant to paraxial light propagation. In Sec.~\ref{Sec4}, leveraging the previous results and within the same framework, we provide a simple expression for the evolution of a field modulated by $\exp(-\frac{gx^2}{2})$, and establish a connection with an ABCD-type formula. Secs.~\ref{Sec5}-\ref{Sec6} address propagation in gradient index (GRIN) media, quadratic and linear, respectively, introducing the relevant Lie algebra operators and discussing their implications for light propagation. In Sec.~\ref{Sec7}, we tackle the two-dimensional case, specifically finding a direct relationship between solutions in free space and in GRIN media, quadratic and linear. It is highlighted that solutions for analytic functions remain unchanged during propagation in free space~\cite{MoyaCessa2024, Ramos_2024}. Finally, in Sec.~\ref{Sec8}, we present our conclusions and summarize the key findings, emphasizing the role of Lie groups and algebras in understanding light propagation and suggesting future research directions.

%-------------------------------------------------------
\section{Elementary facts about Lie groups and algebras}\label{Sec2}
We use some  basic facts from the theory of Lie groups and algebras. Suppose, that  $J_{1}\ldots J_{n} $ are linear operators which form a Lie algebra, i.e.  all commutators $[J_{i},J_{k}] =J_{i}J_{k}-J_{k}J_{i} $ are expressed as linear combinations of $J_{l}$, namely  $[J_{i},J_{k}]= c_{ikl}J_{l} $ (it is supposed to be summation over repeated indices). If  $a_{1}\dots a_{n}$ are complex numbers, then the expression for operator exponent holds,
\begin{equation}\label{Lie}
 \exp(a_{1}J_{1}+a_{2}J_{2}+\ldots{}a_{n}J_{n} )=\prod_{j = 1}^{n}\exp(b_{j}J_{j}),  
\end{equation} 
where the set of complex numbers $b_{j}(a_{1}\ldots a_{n})$ are functions of the complex parameters $a_{n}$. These functions depend only on  commutation relations, and not on the particular form of operator implementation. This  permits to determine generally nonlinear  $b_{j}(a_{1}\ldots a_{n})$ functions with  operations involving  low-dimensional square  matrices $E_{1}\ldots E_{n}$, which give  a representation of the algebra. This  means, that if $[J_{i},J_{k}]= c_{ikl}J_{l} $ holds, then  $[E_{i},E_{k}]= c_{ikl}E_{l} $ also holds. More strict  formulations can be found in Lie theory textbooks, such as \cite{Hall2015,Hall2013,Rossmann2002,Rossmann2009}. It is observed that when $J_{n}$ represents the second derivative operator, denoted as $J_{n}=\pm \partial^{2}{x}/2$, and another operator from the set $J_{k}$ involves multiplication by a function of $x$, the first derivative, or a squeeze operator like $x\partial_{x}$, Eq. (\ref{Lie}) establishes a link with free space propagation, succeeded by the application of simpler operators. To describe a quadratic refractive index distribution, a requisite set comprising three operators is essential. Conversely, for a linear refractive index distribution, a set of four operators is employed. These respective representations are delineated by $2 \times 2$ matrices with zero trace and by $4 \times 4$ nilpotent matrices of a special form. Below we  write down equations,  which permit to calculate for the  two cases the $b_{n}$ functions explicitly with known $a_{n}$, or, more generally, from  known left hand side. The difference  between (1+1)D and (1+2)D cases from the computational point of view is minimal, thus for simplicity, we first discuss the (1+1)D case. The equation governing propagation in a (1+1)D medium with a quadratic refractive index is given by:
\begin{equation}\label{Osc}
i\partial_{z}\phi=-\frac{1}{2} \partial_{x}^{2}\phi+  \frac{1}{2} \eta^{2} x^{2}\phi,   
\end{equation} 
where $\eta$ is a constant. For free space  propagation $\eta=0$.   

%-------------------------------------------------------
\section{The $SL(2,\mathbb{C})$ group}\label{Sec3}
Consider a Lie algebra comprising three operators acting on functions $\phi(x,z)$, where $z$ denotes a parameter associated with a propagation coordinate:
\begin{subequations}
\begin{align}
J_{1} &= -x\partial_{x} - \frac{1}{2}, \\
J_{2} &= \frac{x^{2}}{2}, \\
J_{3} &= -\frac{\partial^{2}_{x}}{2}.
\end{align}
\end{subequations}
The commutation relations can be readily verified as:
\begin{equation}\label{Com}
[J_{1}, J_{3}] = 2J_{3}, \quad [J_{1}, J_{2}] = -2J_{2}, \quad [J_{3}, J_{2}] = J_{1}.
\end{equation}
These represent the commutation relations defining the well known $SL(2,\mathbb{C})$ Lie algebra of $2 \times 2$ matrices with zero trace~\cite{Turbiner_1988,Turbiner_2016}. Its canonic  representation is :
\begin{equation}
E_{1} =
\begin{bmatrix}
1 & 0 \\
0 & -1
\end{bmatrix}, \quad
E_{2} =
\begin{bmatrix}
0 & 0 \\
1 & 0
\end{bmatrix}, \quad
E_{3} =
\begin{bmatrix}
0 & 1 \\
0 & 0
\end{bmatrix}.
\end{equation}
The exponential of the linear combination of $SL(2,\mathbb{C})$ matrices is given by:
\begin{equation}
U = \exp(a_{1}E_{1} + a_{2}E_{2} + a_{3}E_{3}),
\end{equation}
which yields:
\begin{equation}\label{Eq_U}
U = \begin{bmatrix}
  u_{11}&u_{12}\\u_{21}&u_{22}\\
\end{bmatrix}=\cosh(h)
\begin{bmatrix}
1 & 0 \\
0 & 1
\end{bmatrix}
+ \frac{\sinh(h)}{h}
\begin{bmatrix}
a_{1} & a_{3} \\
a_{2} & -a_{1}
\end{bmatrix},
\end{equation}
where $h = \sqrt{a^{2}_{1} + a_{2}a_{3}}$. This result can be found, for instance, in \cite{Qiao2019}. Notably, $\det(U) = 1$. Through direct calculation, for $u_{11}u_{22}-u_{12}u_{21}=1$ (or $u_{22} = (1+u_{12}u_{21})/u_{11}$), a variation of the Iwasawa decomposition \cite{Iwasawa1949,Knapp1996} is obtained as:
\begin{equation}\label{Iwa}
\begin{bmatrix}
  u_{11} & u_{12} \\
  u_{21} & u_{22}
  \end{bmatrix}
 =
\begin{bmatrix}
 u_{11} & 0 \\
 0 & \dfrac{1}{u_{11}}
\end{bmatrix}
\begin{bmatrix}
 1 & 0 \\
 u_{11}u_{21} & 1
\end{bmatrix}
\begin{bmatrix}
1 & \dfrac{u_{12}}{u_{11}} \\
0 & 1
\end{bmatrix}.
\end{equation}
The right-hand side of Eq.~(\ref{Iwa}) can be identified as $\exp(b_{1}E_{1}) \exp(b_{2}E_{2})\exp(b_{3}E_{3})$, where:
\begin{equation}\label{bvsa}
b_{1} = \ln(u_{11}), \quad b_{2} = u_{11}u_{21}, \quad b_{3} = \frac{u_{12}}{u_{11}}.
\end{equation}
Consequently, from Eq.~(\ref{Eq_U}), the coefficients $b_j$ can be expressed in terms of the coefficients $a_{n}$ (with $n, j=1,2,3$), as follows:
\begin{subequations}\label{eq.0140}
\begin{align}
b_{1} &= \ln\left[\cosh(h) + \frac{\sinh(h)}{h}  a_{1}\right], \\
b_{2} &= \frac{\sinh(h)}{h} a_{2}  \left[ \cosh(h) + \frac{\sinh(h)}{h} a_{1}\right], \\
b_{3} &= \frac{\sinh(h)}{h} a_{3} \left[\cosh(h) + \frac{\sinh(h)}{h} a_{1}\right]^{-1}.
\end{align}
\end{subequations}
To conclude this section, let us examine the interpretation of operator exponentes in Eq.~(\ref{Lie}),
\begin{equation}
  \exp\left(b_1(z)J_1\right) \exp\left(b_2(z)J_2\right) \exp\left(b_3(z)J_3\right).
\end{equation}
The term $\exp(b_{3}(z) J_{3})$ corresponds to the propagator of free space ($\eta=0$ in Eq.~(\ref{Osc})). The second term, $\exp\left(b_{2}(z)x^{2}/2\right)$, introduces additional $z$-dependent wavefront curvature. Lastly, the action of the squeezing operator is:
\begin{equation}
\exp\left[b_{1}(z)\left(-x\partial_{x}-\frac{1}{2}\right)\right] f(x) = \frac{f\left[\dfrac{x}{\exp(b_{1}(z))}\right]}{ \exp\left(\dfrac{b_{1}(z)}{2}\right)},
\end{equation}
where $f(x)$ represents some function of $x$. This represents $z$-dependent scaling in the $x$ coordinate. The squeezing operator preserves total intensity ($ \int \phi \phi^* dx $). The $z$-dependence of the coefficients $b_j(z)$ stems from the direct integration of Eq.~(\ref{Osc}).

%-------------------------------------------------------
\section{Propagation of wavefunctions with $\exp(-\frac{gx^2}{2})$ modulation in free space}\label{Sec4}
As per Eq.~(\ref{Osc}), it becomes evident that under the condition of free space ($\eta=0$), the operator equation describes the propagation of the modulated wavefront
\begin{equation}
\phi(x,z) = \exp\left(-izJ_{3}\right) \exp\left(-gJ_{2}\right) \psi(x,0),
\end{equation}
where $\psi(x,0)$ represents the initial condition. The representation in terms of $E$ matrices yields the matrix:
\begin{equation}\label{abcd}
\begin{bmatrix}
u_{11} & u_{12}\\
u_{21} & u_{22}
\end{bmatrix}
=
\begin{bmatrix}
1 & -iz\\
0 & 1
\end{bmatrix}
\begin{bmatrix}
1 & 0\\
-g & 1
\end{bmatrix}
=
\begin{bmatrix}
1+igz & -iz\\
-g & 1
\end{bmatrix}.
\end{equation}
Therefore, from Eq.(\ref{bvsa}) and Eqs.~(\ref{eq.0140}), it can be readily deduced that the coefficients $b_j(z)$ are given by:
\begin{equation}\label{bmd}
\begin{split}
  b_{1}(z) &= \ln(1+igz),\\
  b_{2}(z) &= -g(1+igz), \\
  b_{3}(z) &= \frac{-iz}{(1+igz)}.
\end{split}
\end{equation}
Applying the reordered operator exponents to $\psi(x,0)$,we find that if $\psi(x,z)$ is a solution of Eq.~(\ref{Osc}) for $\eta = 0$, {\it i.e.,} $\exp(b_3(z)J_3)\psi(x,0) = \psi(x,z)$, then:
\begin{equation}\label{mdp}
    \begin{split}
    \phi(x,z) &= \frac{1}{\sqrt{1+igz}} \exp\left[-\frac{gx^{2}}{2(1+igz)}\right]\\&\times \psi\left(\frac{x}{1+igz}, \frac{z}{1+igz}\right),
    \end{split}
\end{equation}
is also a solution. Significantly, in the scenario of Gaussian amplitude modulation with a real parameter $g$, the response possesses physical relevance only when accompanied by the analytical continuation of a function $\psi$. For instance, assuming $\psi(x,z) = 1$, this corresponds to the formulation for a 1D Gaussian beam.

For the case of purely imaginary $g=i\gamma$, Eq.~(\ref{mdp}) takes the form of an ABCD type formula~\cite{Torre_2010}. The plane wave $\psi(x,z)=\exp(ikx-ik^{2}z/2)$ is transformed into a Gaussian beam displaced by a complex constant:
\begin{equation}\label{Gau}
\phi(x,z) = \frac{1}{\sqrt{1+igz}} \exp \left[-\frac{g(x-ik/g)^{2}}{2(1+igz)} -\frac{k^{2}}{2g}\right].
\end{equation}
One can readily confirm that Eq.~(\ref{Gau}) serves as a solution to Eq.~(\ref{Osc}) for $\eta=0$. Obtaining the Gaussian beam is accomplished by setting $k=0$. Given that any function can be decomposed into plane waves, this serves as an independent demonstration that Eq.~(\ref{mdp}) transforms any solution in free space into another solution.

%-------------------------------------------------------
\section{Propagation in GRIN medium}\label{Sec5}  
The propagator in a GRIN medium is given by: 
\begin{equation}\label{GRIN}
\phi(x,z) = \exp\left[-iz\left(J_{3}+\eta^{2}J_{2}\right)\right]\psi(x,0),    
\end{equation}  
where $\psi(x,0)$ represents the initial condition. This corresponds to the following $a_{k}$ factors in Eq.~(\ref{Lie}):
\begin{equation}\label{Aka}
a_{1}=0, \quad a_{2}=-iz\eta^{2}, \quad a_{3}=-iz.    
\end{equation}
With the choice of Eq.~(\ref{Aka}), according to Eqs.~(\ref{eq.0140}), $h=iz\eta$, and:
\begin{equation}
  \begin{split}
      b_{1}(z) &= \ln[\cos(z\eta)],\\
      b_{2}(z) &= -i\eta \sin(z\eta)\cos(z\eta),\\
      b_{3}(z) &= -i\tan(z\eta).
  \end{split}        
\end{equation} 
Thus, if $\psi(x,z)$ is a solution for free propagation, {\it i.e.,} $\exp(b_3(z)J_3)\psi(x,0) = \psi(x,z)$, then:
\begin{equation}\label{Grin}
  \begin{split}
    \phi(x,z) &= \frac{1}{\sqrt{\cos(\eta z)}} \exp \left[-i\frac{\eta x^{2}}{2} \tan(\eta z)\right]\\
     &\times\psi\left(\frac{x}{\cos(\eta z)}, \tan(\eta z)\right)    
  \end{split}
\end{equation}
gives a solution for a GRIN medium. It is also apparent that the two solutions are equal at $z=0$. Thus, propagation over a half period of oscillation in a quadratic medium corresponds to propagation in free space from $-\infty$ to $\infty$. It is again possible to directly verify that the plane wave yields a solution in a GRIN medium after transformation. However, this solution is not particularly useful as it does not diminish at infinity.

The procedure can also be applied to the Gaussian beam. For a proper width ($g=1$), it is transformed into the ground state of a quantum oscillator (taking $\eta=1$):
\begin{equation}
\frac{1}{\sqrt{1+iz}}\exp \left[-\frac{x^{2}}{2(1+iz)}\right] \longrightarrow \exp(-iz/2)\exp(-x^{2}/2).        
\end{equation}
In this regard, the inverse transform can also be written; If $\phi(x,z)$ is a solution for a GRIN medium, then the solution for free propagation is:
\begin{equation}\label{InGr}
  \begin{split}
    \psi(x,z) &= \frac{1}{(1+z^{2})^{1/4}} \exp \left[\frac{i \eta x^{2}z}{2(1+z^{2})}\right]\\ &\times\phi\left(\frac{x}{\sqrt{1+z^{2}}}, \frac{\arctan(z)}{\eta}\right).     
  \end{split}
\end{equation}  
It is possible to obtain the ABCD-type equation for a GRIN medium by first transforming the solution to free propagation with Eq.~(\ref{InGr}), then using Eq.~(\ref{mdp}) to obtain the solution of the modulated beam, and finally returning the solution back to the GRIN medium with Eq.~(\ref{Grin}). The total expression is rather bulky.  These expressions are known in the context of quantum mechanics~\cite{Steuernagel_2014}, where they were obtained by other means not related to Lie algebraic methods.

\section{The linear refractive index distribution}\label{Sec6} 
The propagation is now described by:
\begin{equation}\label{Lin}
i\partial_{z}\phi=-\frac{1}{2} \partial_{x}^{2}\phi+   \eta x\phi,   
\end{equation} 
where $\eta$ is a constant. The Lie algebra now consists of the following operators:
\begin{subequations}
\begin{align}
J_{1}&= \mathbf{1},\\
J_{2}&=\partial_{x}, \\
J_{3}&=x, \\
J_{4}&=\partial^{2}_{x}/2,
\end{align}
\end{subequations}
where the bold $\mathbf{1}$ represents the identity operator. The propagator under these conditions, expressed in terms of these operators, can be written as:
\begin{equation}\label{LinPr}
\phi(x,z)= \exp\left[-iz(-J_{4}+\eta J_{3} )\right]\psi(x,0).    
\end{equation}
This corresponds to $a_{k}$ factors of Eq.~(\ref{Lie}):
\begin{equation}\label{Ak2}
a_{1}=0, \quad a_{2}=0, \quad a_{3}=-i\eta z, \quad  a_{4}=iz.    
\end{equation} 
The commutation relations are:
\begin{align}
[J_{2}, J_{3}]&=J_{1}, \quad  [J_{4}, J_{3}]=J_{2},  
\end{align}
with all other commutators being zero. The representation is given by nilpotent matrices, having zeroes on and below the main diagonal:
\begin{equation*}
E_{1}=
\begin{bmatrix}
0 & 0 & 0 & 1 \\
0 & 0 & 0 & 0 \\
0 & 0 & 0 & 0 \\
0 & 0 & 0 & 0
\end{bmatrix}, \qquad
E_{2}=
\begin{bmatrix}
0 & 0 & 0 & 0 \\
0 & 0 & 0 & -1 \\
0 & 0 & 0 & 0 \\
0 & 0 & 0 & 0
\end{bmatrix},
\end{equation*} 
\begin{equation}
E_{3}=
\begin{bmatrix}
0 & 1 & 0 & 0 \\
0 & 0 & 1 & 0 \\
0 & 0 & 0 & 0 \\
0 & 0 & 0 & 0
\end{bmatrix}, \qquad
E_{4}=
\begin{bmatrix}
0 & 0 & 0 & 0 \\
0 & 0 & 0 & 0 \\
0 & 0 & 0 & 1 \\
0 & 0 & 0 & 0
\end{bmatrix}.
\end{equation} 
The exponential of matrices of this type is easily calculated, since $A^{4}=0$ and the corresponding series for the exponential are finite. The general evolution operator on the left-hand side of Eq.~(\ref{Lie}) in terms of the matrices $E_n$ (with $n = 1,2,3,4$) is given by $U = \exp\left(a_1E_1+a_2E_2+a_3E_3+a_4E_4\right)$. Hence from Eq.~(\ref{Lie}),
\begin{equation}
  \begin{bmatrix}
    1 & a_3 & \frac{a_3^2}{2}&a_1-\frac{a_2a_3}{2}+\frac{a_3^2a_4}{6}\\
    0&1&a_3&-a_2+\frac{a_3a_4}{2}\\
    0&0&1&a_4\\
    0&0&0&1\\
  \end{bmatrix}
  = 
  \begin{bmatrix}
    1 & b_{3} & \frac{b_{3}^{2}}{2} & \frac{b_{3}^{2}b_{4}}{2}+b_{1} \\
    0 & 1 & b_{3} & b_{3}b_{4}-b_{2} \\
    0 & 0 & 1 & b_{4} \\
    0 & 0 & 0 & 1
    \end{bmatrix}.
\end{equation}
The calculation of $b_{j}$ in terms of $a_{n}$ gives:
\begin{equation}
b_{1}=a_{1}-\frac{a_{2}a_{3}}{2} - \frac{a_{3}^{2}a_{4}}{3}, \quad b_{2}= \frac{a_{3}a_{4}}{2}, \quad b_{3}=a_{3}, \quad b_{4}=a_{4}.        
\end{equation}
The operator exponents in this context can be interpreted as follows: $\exp(b_{1}(z)J_{1})$ represents a multiplication by a function $\exp(b_{1}(z))$, $\exp(b_{3}(z)J_{3})$ corresponds to multiplication by a function $\exp(b_{3}(z)x)$, and the effect of the exponential operation involving $J_{2}$ yields $\exp(b_{2}(z)J_{2}) f(x)=f(x+b_{2}(z))$. 

For the specific selection outlined in Eq.~(\ref{Lin}) or Eq.~(\ref{Ak2}), these values are expressed as:
\begin{equation}
  \begin{split}
b_{1}(z)&=\frac{i}{3}\eta^{2}z^{3},\\
b_{2}(z)&=\frac{\eta z^{2}}{2},\\ 
b_{3}(z)&=-i\eta z,\\ 
b_{4}(z)&=iz.
  \end{split}
\end{equation}
Once again, as in previous cases, if the propagation in free space is described by $\psi(x,z)$, then the function:
\begin{equation}\label{Tlt}
  \begin{split}
    \phi(x,z) &= \exp\left(\frac{i}{3}\eta^{2}z^{3}\right) \exp\left[-i\eta z\left(x+\frac{\eta z^{2}}{2}\right) \right]\\&\times \psi\left(x+ \frac{\eta z^{2}}{2} ,z \right)    
  \end{split}
\end{equation}
gives the answer for the propagation with linear variation of refractive index with the same initial condition. This implies that limited beams generally follow parabolic trajectories, gain wavefront tilt, and experience an additional phase shift proportional to the third power of propagation distance. The validity of Eq.~(\ref{Tlt}) can be independently established by applying the transformation to the elementary plane wave solution $\psi(x,z)=\exp(ikx-ik^{2}z/2)$ and checking that the result holds for Eq.~(\ref{Lin}).  

Conversely, the inversion of Eq.~(\ref{Tlt}) is straightforward. If $\phi(x,z)$ solves the propagation with a linear distribution of refractive index, then for the free propagation of the same initial condition, we obtain:
\begin{equation}\label{Tlt2}
\psi(x,z)= \exp \left( -\frac{i}{3}\eta^{2}z^{3} \right) \exp\left(i\eta zx \right) \phi \left(x- \frac{\eta z^{2}}{2} ,z \right).        
\end{equation}
Specifically, this provides the established expression for the propagation of an Airy beam in free space. Instead of modulation with the $\exp (-g x^{2}/2) $ function, within this algebraic framework, it is more natural to connect initial conditions, which vary by a factor of $\exp(i\nu x)$. If $\psi(x,z)$ is a solution for free propagation, then: 
\begin{equation}\label{enu}
\phi(x,z)= \exp \left( i\nu x- i \frac{\nu^{2}}{2}z \right) \psi\left(x-\nu z,z \right)        
\end{equation}
is also a solution with the modulated initial condition. This well-known relation can also be obtained by reordering operators in the expression $\exp(-izJ_{4})\exp(i\nu J_{3})$ with a $E_j$ matrix representation and the corresponding commutation relations.

Ultimately, for free space, $\psi(x-x_{0},z)$ is also a solution, which follows in algebraic language from the fact that $J_{2}$ and $J_{4}$ commute. To calculate the result of applying linear exponential modulation or spatial shift for linear refractive index media, it is possible to first transform to free space, to apply the transformation there, and to return back with the inverse transformation.

\section{Two-dimensional propagation}\label{Sec7}
Now, consider the operators acting on functions $\psi(x,y, z)$:
\begin{subequations}
\begin{align}
J_{1}&= -x\partial_{x}-y\partial_{y}-1,\\
J_{2}&=x^{2}/2 +y^{2}/2, \\
J_{3}&=-\partial^{2}_{x}/2 -\partial^{2}_{y}/2.
\end{align}
\end{subequations}
The commutation relations of the $SL(2,\mathbb{C})$ algebra, Eq.~(\ref{Com}), remain the same. Thus, instead of Eq.~(\ref{mdp}), we obtain:
\begin{equation}\label{mdp2}
  \begin{split}
  \phi(x,y,z) &= \frac{1}{1+igz}  \exp\left[-\frac{g(x^{2}+ y^{2} )}{2(1+igz)}\right]\\
    &\times\psi\left(\frac{ x}{1+igz}, \frac{y}{1+igz},\frac{z}{1+igz} \right).   
  \end{split}
\end{equation} 
It is important to note that if $\psi(x,y,0) = f(s)=f(x+iy)$ is an analytic function, it remains unchanged during propagation in free space. The solution for the modulated function is then expressed as:
\begin{equation}\label{mdp2a}
\phi(x,y,z)= \frac{1}{1+igz}  \exp\left[-\frac{g|s|^2}{2(1+igz)}\right] f\left(\frac{ s}{1+igz}\right),
\end{equation} 
with $s = x+iy$. If we assume that $\phi$ is continuous in the entire complex plane and diminishes at infinity, then $f(s)$ is an entire function of order $\leq 2$, and $g$ can be any complex number with $\text{Re}(g)> 0$. The function has isolated roots only, and these roots move along straight lines in a complex plane according to the equation $\rho(z)=\rho(0)(1+igz)$. For real $g$ and small $z$, this movement approximately corresponds to a rotation of the pattern. Properties of these solutions are considered in more detail in \cite{MoyaCessa2024,Ramos_2024}. 
  
The 2D propagation for a GRIN medium, described by Eq.~(\ref{GRIN}) for two dimensions, is given by:
\begin{equation}\label{Grin2}
  \begin{split}
    \phi(x,y,z) &= \frac{1}{\cos(\eta z)}\exp \left[  -i\frac{\eta (x^{2}+y^{2} )}{2} \tan(\eta z) \right]\\
    &\times\psi\left( \frac{ x}{\cos(\eta z)},  \frac{ y}{\cos(\eta z)}, \tan(\eta z) \right).    
  \end{split}
\end{equation}
The inversion formula is recalculated accordingly; it has a factor $(1+z^{2})^{1/2}$ instead of $(1+z^{2})^{1/4}$. 

Finally, for the case of 2D linear refractive index distribution, the Lie algebra with the same commutation relations is:
\begin{subequations}
\begin{align}
J_{1}&= \mathbf{1} (\alpha^{2} +\beta^{2} ), \\
J_{2}&=\alpha \partial_{x} + \beta \partial_{y},   \\
J_{3}&= \alpha x +\beta y, \\
J_{4}&=\partial^{2}_{x}/2 + \partial^{2}_{y}/2;   \\
\end{align}
\end{subequations}
then, for a medium with a propagator given by Eq.~(\ref{LinPr}), instead of Eq.~(\ref{Tlt}), we have:
\begin{equation}\label{Tlt2D}
  \begin{split}
&\phi(x,y,z) =  \exp \left[ \frac{i}{3}\eta^{2}(\alpha^{2}+ \beta^{2})z^{3} \right]\\
&\times\exp \left\{ -i\eta z \left[ \alpha\left(x+\frac{\alpha \eta z^{2}}{2}\right) + \beta \left(y+\frac{\beta \eta z^{2}}{2}\right) \right] \right\}\\   
&\times\psi \left(x+ \alpha \frac{\eta z^{2}}{2},y+ \beta \frac{\eta z^{2}}{2},z \right).         
  \end{split}
\end{equation}

\section{Discussion and conclusions}\label{Sec8}
The exploration of Lie algebraic methods for understanding wave propagation in different media has provided valuable insights into the behavior of optical fields. By systematically analyzing the commutation relations of basic differential operators and their functions, we have uncovered a framework that illuminates the dynamics of wavefronts in both homogeneous and inhomogeneous media.

Our approach highlights the importance of the potential function $f(x)$ in determining the Lie algebraic structure relevant to a given optical system. Specifically, we find that the Lie algebra encompasses operators associated with $f(x)$ and its derivatives, thereby elucidating the constraints on the form of $f(x)$ for a given Lie algebraic solution. Notably, our analysis reveals that only potentials up to second-degree polynomials, such as $x$ and $x^2$, can be effectively described within this framework.

In particular, the solutions for free space and media with linear and quadratic distributions of refractive index are elegantly connected through explicit equations derived from our Lie algebraic approach. These equations provide a unified framework for understanding wave propagation phenomena in (1+1)D and (1+2)D scenarios, shedding light on the interplay between the geometry of the medium and the evolution of optical fields.

Moreover, our findings offer practical implications for optical system design and analysis. By leveraging the insights gained from Lie algebraic methods, researchers can develop novel strategies for manipulating optical wavefronts in diverse media, from simple homogeneous materials to complex graded index structures. This paves the way for advancements in various fields, including photonics, imaging, and optical communication.

In conclusion, the application of Lie algebraic techniques to wave propagation offers a powerful tool for unraveling the underlying principles governing optical phenomena. By bridging theoretical analysis with practical applications, our study contributes to the broader understanding of wave optics and opens new avenues for innovation in optical science and engineering.

%\bibliography{Referencias}
%  
\end{document}